# Monolayer, Bilayer and Heterostructures of Green Phosphorene for Water Splitting and Photovoltaics

Sumandeep Kaur[1, 2], Ashok Kumar[2*], Sunita Srivastava[1,3*], K. Tankeshwar[1, 3*] and Ravindra Pandey[4]

[1]*Department of Physics, Panjab University, Chandigarh 160014, India*
[2]*Department of Physical Sciences, School of Basic and Applied Sciences, Central University of Punjab, Bathinda, 151001, India*
[3]*Department of Physics, Guru Jambheshwar University of Science and Technology, Hisar, 125001, Haryana, India*
[4]*Department of Physics, Michigan Technological University, Houghton, MI, 49931, USA*

(October 19, 2018)

*Emails:

Ashok Kumar (ashokphy@cup.edu.in)

Sunita Srivastava (sunita@pu.ac.in )

K. Tankeshwar (tkumar@gjust.org)

# Abstract

We report the results of density functional theory (DFT) based calculations on monolayer and bilayer green phosphorene and their heterostructures with MoSe$_2$. Both monolayer and bilayer green phosphorene are direct band gap semiconductors and possess anisotropic carrier mobility as high as $10^4$ cm$^2$V$^{-1}$s$^{-1}$. In bilayers, pressure of about 9 GPa induces the semiconductor-metal transition. Moreover, the band gap depends strongly on the thickness of the films and the external electric field. By employing strain-engineering under suitable solution conditions, monolayer and AC-stacked bilayer green phosphorene offer the band edge alignments which can be used for water splitting. The upper limit of the power conversion efficiencies for monolayer, AB- and AC-stacked bilayer green phosphorene heterostructures with MoSe$_2$ is calculated to be 18-21 %. Our results show the possibility of green phosphorene to be used as photocatalytic and photovoltaic material in the energy-related applications.

**Keywords:** Density Functional Theory, Green Phosphorene, Carrier Mobility, Photovoltaics, Water Splitting.

# 1. Introduction

After the successful realization of free standing black phosphorene [1, 2], there exists growing interest among scientific community to explore about its isostructural allotropes. Compared to the previously known 2D materials, phosphorene shows great potential to be used in practical applications due to its unique properties [3-8] viz structural anisotropy, high carrier mobility (~$10^3$ cm$^2$V$^{-1}$s$^{-1}$), high lateral flexibility and tunability of its bandgap with number of layers, stacking pattern, electric field and strain. Due to the inequivalent sp$^3$ orbital hybridization in its structure, phosphorous in its bulk form is known to exhibit various allotropes like red, white, violet and black [3]. Since 2014, various allotropes of phosphorene consisting of honeycomb structures (α-, β-, γ-, δ- and red-phosphorene) and non-honeycomb structures (ε-, ς-, η-, θ-, octa-, hexstar- and ψ-phosphorene) have been predicted [9-17]. All these allotropes are semiconducting with bandgap ranging from 0.48 eV for ε-phosphorene to 2.09 eV for octa-phosphorene [17]. Out of all these allotropes, only α- and β-phosphorene have been experimentally realized [18-21].

The very recently predicted allotrope of phosphorene is green phosphorene [22, 23] which is constructed from the black phosphorene by flipping every twelfth row of bonds upside down followed by the dislocation of armchair ridges after every forth row. Green phosphorene possesses monoclinic C$_2$/m structure with six atoms per unit cell. It shows strong anisotropy in its transport and optical properties. It has been predicted to be more stable than β-P and can be formed from α-P at temperatures above 87 K. Also, it is predicted to be synthesised on corrugated metal substrates [22].

Motivated by the recent study on green phosphorene, we have investigated bilayer green phosphorene in two types of stacking pattern by applying perpendicular electric field, vertical pressure and in-plane uniaxial and biaxial strain. By employing the deformation potential theory, the carrier mobilities of the monolayers and bilayers are investigated. The possibility of strain engineered monolayer and bilayers green-phosphorene to be used for water splitting photocatalytic application has been explored. From application point of view, we combine the monolayer and bilayers of green phosphorene with $MoSe_2$ to form heterostructures which comes out to be an appropriate combination to be used as a solar cell. The calculated upper limits to the power conversion efficiencies of heterostructure systems are found to be comparable to the black phosphorene/$MoS_2$ based heterostructures [24, 25].

## 2. Computational Details

SIESTA simulation package has been employed to perform all the calculations [26]. Norm- conserving Troullier Martin pseudopotential in fully separable Kleinman and Bylander form is used to treat the electron-ion interactions [27]. The exchange and correlation energies are treated within the van der Waals (vdW)-DRSLL functional [28]. The Kohn-Sham orbitals were expanded as a linear combination of numerical pseudo atomic orbitals using a split-valence double zeta basis set with polarization functions (DZP). Throughout geometry optimization, the confinement energy of numerical pseudo-atomic orbitals is taken as 0.01 Ry. Minimization of energy was carried out using standard conjugate-gradient (CG) technique. Structures were relaxed until the forces on each atom were less than 0.01 eV/Å. Monkhorst-Pack scheme is used to sample Brillouin zone with a 30×30×1 mesh for all the structures. The spacing of the real space used to calculate the Hartree exchange and correlation contribution of the total energy was 450 Ry. A vacuum region of about 20 Å perpendicular to 2D plane is used in calculations to prevent the superficial interactions between the periodic images.

## 3. Results and Discussions
### 3.1 Monolayer and Bilayer Green Phosphorene

The bulk green phosphorus exists in monoclinic structure with space group C2/m. The calculated optimized lattice parameters are a = 10.93 Å, b = 3.34 Å, c = 8.26 Å, $\alpha = 90.00^0$, $\beta = 56.36^0$, $\gamma = 90.00^0$ [Figures S1], which are in very good agreements with the previously reported values [22]. The calculated direct band gap is 0.32 eV at the GGA-PBE level of theory. Note that calculated $G_0W_0$ quasiparticle band gap is 0.68 eV [22]. This dissimilarity in bandgap value is due to the well-known fact that GGA underestimates the bandgaps of the materials. In order to describe interlayer van der Waals interactions correctly, we include vdw-DRSLL functional for the calculations of multilayer systems.

**3.1 Structures and Energetics**

It has been established that the AB stacking is the energetically preferred in 2D materials. In bilayer black phosphorene, AB stacking is the most favourable followed by AC and AA stacking [24]. This is also the case with green phosphorene for which AB stacking is preferred over AA stacking. Therefore, we have considered two most energetically preferred stacking pattern in our study. Figures 1 and Figure S2 illustrate structures of the monolayer, AB- and AC-stacked bilayers of green phosphorene in the rectangular unit cells. AB-stacking is formed by shifting one of the two layers by half the cell either along x or y direction. For AC-stacking, one of the layers is a mirror image of the other layer. The in-plane and out of plane bond lengths in monolayer green phosphorene are not affected much when bilayers are formed. The interlayer distance of AB- and AC-stacked bilayer green phosphorene is calculated to be 3.10 Å and 3.75 Å, respectively [Table 1].

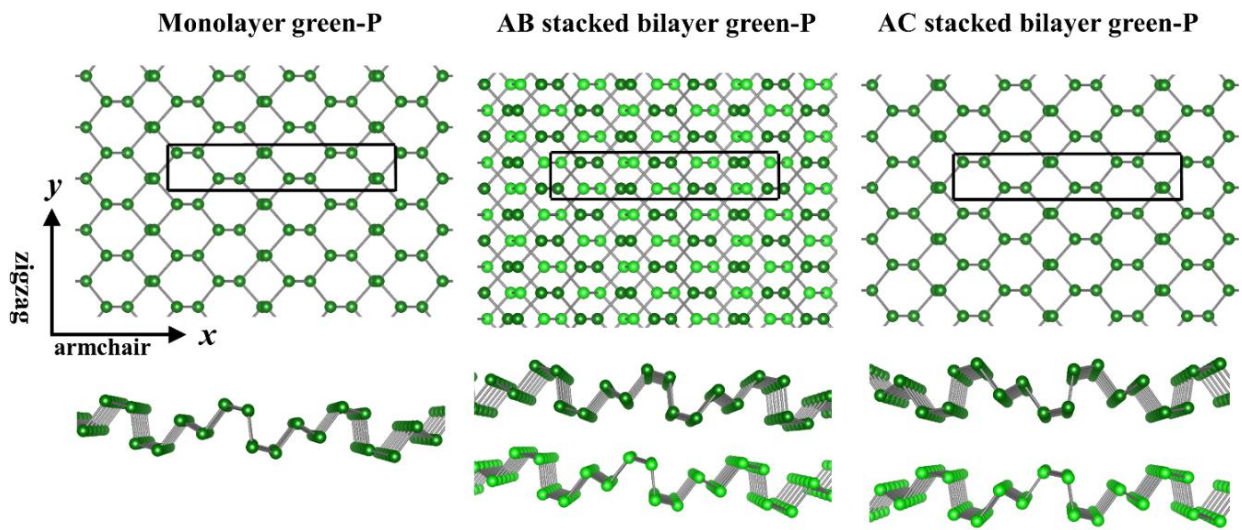

**FIGURE 1:** Top and side views of structures of monolayer green phosphorene, AB-stacked bilayer and AC-stacked bilayer green phosphorene. Rectangular box represents the unit cell.

**TABLE 1.** Green phosphorene**:** Lattice constant (*a, b*), number of atoms per unit cell (N), cohesive energy ($E_C$), elastic modulus (*C*) of monolayer and AB- and AC- stacked bilayer green-P. The interlayer distance (*d*) and binding energy ($E_B$) of bilayers is also given.

| Systems | Monolayer | AB-stacked bilayer | AC-stacked bilayer |
|---|---|---|---|
| *a, b* (Å) | 14.83, 3.42 | 14.87, 3.42 | 14.89, 3.42 |
| $E_C$ (eV/atom) | 4.97 | 5.00 | 5.02 |
| No of Atoms | 12 | 24 | 24 |
| $C_{x\_2D}$ (N/m) | 26.5 | 54.5 | 53.6 |
| $C_{y\_2D}$ (N/m) | 84.9 | 168.5 | 170.9 |
| *d* (Å) | --- | 3.14 | 3.75 |
| $E_B$ (eV/atom) | --- | 0.52 | 0.26 |

In order to determine the energetic stability, we calculate the cohesive energy which is defined as the amount of energy required to break the monolayer or bilayer green phosphorene into isolated single phosphorous atoms, i.e., $E_C = \frac{E_A - NE_S}{N}$, where $E_A$ is the total energy of the considered monolayer or bilayer green phosphorene and $E_S$ is the energy of isolated single phosphorous atom and *N* is the number of atoms per unit cell of the monolayer or bilayer green phosphorene. The calculated cohesive energy (4.97 eV/atom) of green phosphorene is nearly equal to that of black phosphorene (4.98 eV/atom) which indicate nearly equal stability of both monolayers. The cohesive energy of AB- and AC- bilayers are calculated to be 5.00 eV/atom and 5.02 eV/atom, respectively that depicts the bilayer systems to be energetically more favourable. Also, small binding energy values [Table 1] of the bilayers suggest that the weak van der Waals forces acts between the layers. Note that the binding energy is calculated as: $E_B = \frac{E_A - (E_1 + E_2)}{N}$, where $E_A$ is the total energy bilayer green phosphorene and $E_1$ and $E_2$ are the energies of the individual layers of the bilayer system within the cell of the bilayer and *N* is the number of atoms per unit cell of the bilayer green phosphorene.

### 3.2 Mechanical Properties

In order to investigate the mechanical stability of considered mono and bilayer systems, we calculate the ultimate tensile strength and strain by calculating strain-stress relationship which can be obtained by calculating the components of the stress tensor with respect to the strain tensor [9]. Here, the stress tensor is the positive derivative of the total energy with respect to the strain tensor. The ultimate tensile strength (UTSH) is referred to the maximum stress that a layer can withstand before breaking. The point where the slope of the strain–stress curve becomes zero gives the value of (UTSH) and the maximum strain at this point represent the magnitude of ultimate tensile strain (UTSR).

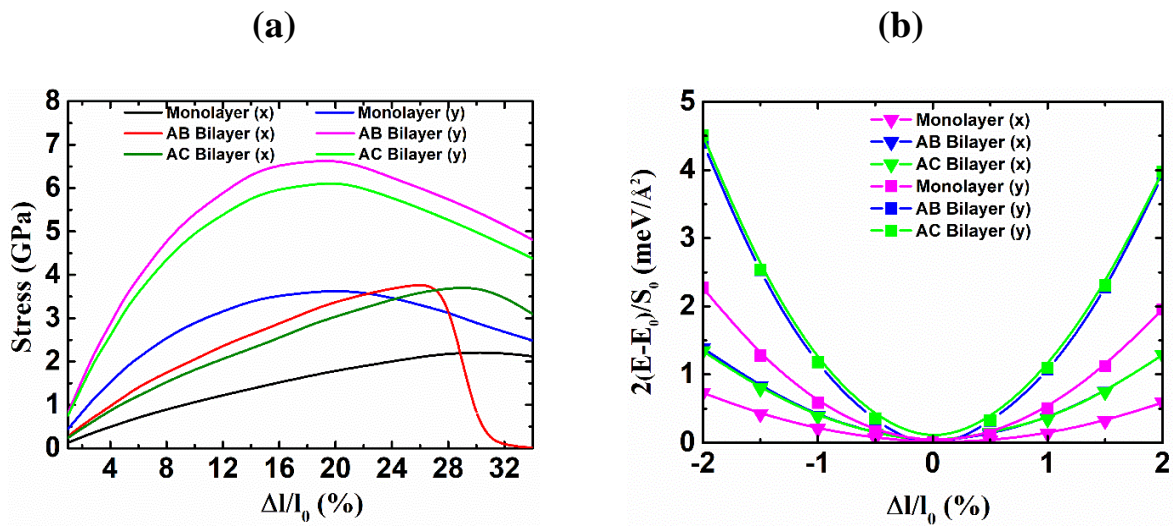

**FIGURE 2:** (a) Strain-stress curves and (b) strain versus strain energy density curve for monolayer, AB- and AC-stacked bilayer of green phosphorene.

Monolayer, AB-stacked bilayer and AC-stacked bilayer green-P can withstand UTSR of up to 30% (20%), 26% (20%) and 30% (20%) along armchair (zigzag) direction, respectively, with the corresponding UTSH of 2.21GPa (3.63GPa), 3.70GPa (6.65GPa) and 3.68GPa (6.13GPa), respectively [Figure 2(a)]. The UTSR values for the monolayer and AC-stacked bilayer green-P are the same. However, the corresponding degree of stress that these layers can with stand at the UTSR value suggest the monolayer to be relatively flexible. G. Yang et al., have also predicted a higher UTSR in monolayer green phosphorene along the armchair direction as compared to the zigzag direction [18]. The calculated values of UTSR for monolayer green-P are comparable with monolayer black-P (30% along armchair and 27% along zigzag) [32], monolayer $MoS_2$ (24%), monolayer $MoSe_2$ (26%) and monolayer $MoTe_2$ (28%) [33], while it is much higher compared to monolayer blue phosphorene (16%) [34].

In order to get additional insight into the mechanical stability, we calculate the elastic modulus by applying strain ranging from -2% to +2% in steps of 0.5% along *x* or *y* direction and quadratic fitting of equation 2. Figure 2(b) shows the strain energy density curve for monolayer, AB stacked bilayer and AC stacked bilayer green phosphorene. It can be seen that for both monolayer and bilayer green-P the energy density along *x* direction is always lower than that along *y* direction which depicts the highly anisotropic nature of mono- and bi-layers green phosphorene. The magnitude of the stiffness however, increases as we move from monolayer to bilayer [Table 2]. Also, the flexibility of the layers is higher along the armchair direction compared to the zigzag direction owing to a low value of inplane stiffness along armchair direction compared to zigzag direction [Table 2], which is in-line with the values reported for other allotropes of phosphorene such as monolayer black-P (24.3 N/m along armchair and 103.3 N/m along zigzag direction) and tricycle type phosphorene (12.4 N/m along armchair and 89.1 N/m along zigzag direction) [7, 14].

### 3.3 Effect of Thickness, Electric Field and Pressure on Band Gap

Monolayer green-P is a semiconductor with a direct bandgap of ~1.35 eV. The direct nature of the bandgap is preserved on moving from monolayer to bilayer but the magnitude of bandgap decreases to 1.17 eV and 1.23 eV for AB-stacked AC-stacked bilayer, respectively. The band gaps of ABAB…. type and ACAC… type stacked layers show a decreasing trend with the increase in thickness of the films due to the quantum confinement effect of the charge carriers. These results are in agreement with the other 2D materials such as black phosphorene and $MoS_2$. [Figure 3(a)] . For ABAB… type stacking the bandgap shows direct to indirect transition on moving from N=3 to N=4 and again a transition from indirect to direct bandgap at N=7. Similarly, for ACAC… stacking a direct to indirect bandgap transition occurs on moving from N=5 to N=6.

It is well known that electric field and pressure can effectively tune the electronic properties of 2D materials [35, 36]. On applying perpendicular electric field in bilayer green phosphorene, a systematic shift in both VBM and CBM has been found [Figure 3(b)]. At a critical value of field i.e. ±1.0V/Å and ±1.2V/Å for AB- and AC-stacked bilayers, respectively, band gap closure occur which may be attributed to the electric field induced charge redistribution in the van der Waals gap in bilayers. On applying vertical pressure, metallization occur for AB- and AC-stacked bilayers at 9 GPa and 8 GPa, respectively [Figure 3(c)], which is feasible in real experimental situation. In general, application of pressure via decreasing the interlayer distance in a bilayer facilitates a higher degree of electrons hopping which, in general, can lead to semiconductor-metal transition in the lattice. Note that pressure on bilayer system is calculated as the energy per unit area that is essential to decrease the inter layer distance by $\Delta R = (R_0-R)$ [9]:

$$P = \frac{E-E_0}{\Delta R \times A} \quad (3)$$

where E-E$_0$ is the change in total energy in moving from strained to equilibrium configurations, A is the area of the unit cell and $\Delta R$ (R-R$_0$) is the difference of the interlayer distance at strained and equilibrium configuration.

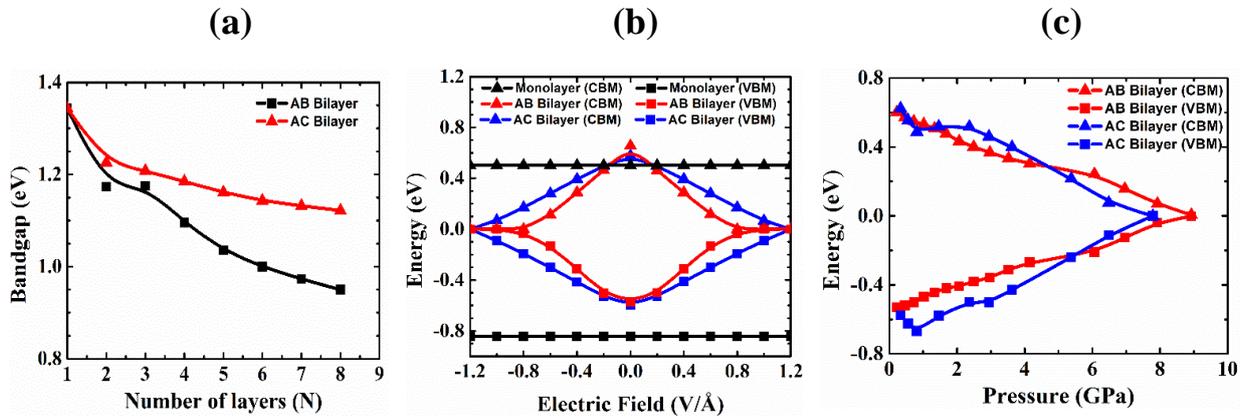

**FIGURE 3:** (a) Variation of bandgap with number of layers (b) variation of VBM and CBM with applied perpendicular electric filed and (c) variation of VBM and CBM variation with vertical pressure. In (b) and (c) zero is Fermi energy.

## 3.4 Carrier Mobilities

We apply a phonon-limited scattering model including the anisotropic characteristics of effective mass, elastic modulus and deformation potential to calculate room temperature (T=300 K) electron and hole mobilities using the formula [29, 30]:

$$\mu_{2D} = \frac{e\hbar^3 C_{2D}}{K_B T m^* m_a^* E_i^2} \quad (1)$$

where, $e$ is the electronic charge, $\hbar$ is the reduced Plank's constant, T is the temperature, $k_B$ is the Boltzmann constant, $C_{2D}$ is in-plane elastic modulus in the propagation direction which is calculated using the formula:

$$(E-E_0)/S_0 = C_{2D}(\Delta l/l_0)^2/2, \quad (2)$$

where *E-E$_0$* represents total energy change, *S$_0$* is the area of the 2D cell and *Δl/l$_0$* is the deformation along x or y direction. $m^*$ is the effective mass in the transport direction (i.e., either along *x* or *y* direction), $m_a^*$ is the average effective mass given by $\sqrt{m_x^* m_y^*}$. $E_i$ in equation (1) is the deformation potential constant calculated using the formula $E_i = \frac{dE_{edge}}{de}$ where $E_{edge}$ is the energy of the CBM (VBM) for electrons (holes) and $e = \Delta l/l_0$ [31].

The carrier mobilities of 2D materials, which mostly governs their electronic properties, depends upon the carrier effective mass, elastic modulus and the deformation potential. It is found that the hole effective mass is highly anisotropic while the electron effective mass remains almost isotropic for monolayer as well as AB- and AC-stacked bilayer green phosphorene. This behaviour is similar to the one reported for black phosphorene in literature [31]. The hole effective mass along zigzag direction is 8.6, 8.4 and 8.0 times that along the armchair direction for monolayer, AB-stacked bilayer and AC stacked bilayer green phosphorene, respectively [Table 2]. Deformation potential (DP) describes the change in energy of the electronic band with elastic deformation. Lower value of deformation potential contributes to an increase in the mobility of electrons or holes. By including the anisotropic character of elastic constants [Table 1], effective mass and deformation potential, we estimate the carrier mobility of monolayer as well as bilayer green phosphorene using equation 1.

The electron mobilities are calculated to be higher than the hole mobilities of the monolayer as well as AB stacked bilayer green-P, indicating their *n*-type behavior. AC- stacked bilayer green-P, on the other hand, behaves *p*-type along armchair direction while it shows *n*-type character along the zigzag direction. The order of magnitude of carrier mobilities (10 -10$^4$ cm$^2$/V/s) reported in our calculations [Table 2] are comparable with black phosphorene [31]. The electrons (holes) mobility anisotropy is calculated to be 19.7 (1.6), 2.3 (1.4) and 64.4 (68.5) for monolayer, AB-stacked bilayer and AC-stacked bilayer green phosphorene, respectively. Note that, the mobility anisotropy is obtained as $R_a = \frac{Max(\mu_x,\mu_y)}{Min(\mu_x,\mu_y)}$, where $R_a = 1$ for isotropic systems and $R_a > 1$ for anisotropic systems [37].

**TABLE 2.** Electron and hole effective mass ($m^*$), deformation potential ($E_1$) and carrier mobility ($\mu$) monolayer green-P, AB- and AC-stacked bilayer green-P along the armchair and zigzag directions.

| Systems | $m^*$ (x) ($m_e$) | $m^*$ (y) ($m_e$) | $E_{1x}$ | $E_{1y}$ | $\mu$ (x) cm$^2$/V/s | $\mu$ (y) cm$^2$/V/s |
|---|---|---|---|---|---|---|
| monolayer (*e*) | 0.27 | 0.20 | -5.02 | 2.35 | 355 | 7020 |
| monolayer (*h*) | 0.21 | 1.71 | -9.93 | 7.71 | 48 | 24 |
| AB bilayer (*e*) | 0.3 | 0.25 | 1.47 | -4.26 | 7226 | 3166 |
| AB bilayer (*h*) | 0.23 | 1.96 | -2.17 | -1.12 | 1589 | 2179 |
| AC bilayer (*e*) | 0.27 | 0.21 | 4.04 | -1.01 | 1104 | 71152 |
| AC bilayer (*h*) | 0.23 | 1.83 | -0.77 | 4.01 | 13159 | 192 |

## 3.5 Strain Engineering in Monolayer and Bilayer Green Phosphorene: Potential Application in Photocatalytic Water Splitting

Considering the high carrier mobility of green phosphorene, we explore the feasibility to use green phosphorene for photocatalytic water splitting. Note that the material to act as photocatalyst for water splitting requires its bandgap to be at least 1.23 eV. Additionally, the conduction band minimum should be more positive than the redox potential of H$_2$/H$^+$ (0V vs NHE) and the valance band maximum should be more negative than the redox potential of H$_2$O/O$_2$ [38].

The overall process of water splitting is as follows:

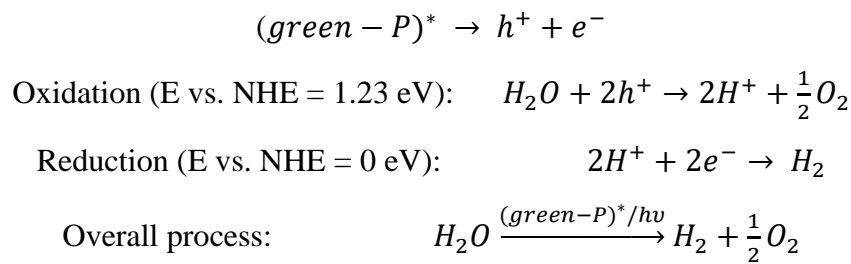

$$(green - P)^* \rightarrow h^+ + e^-$$

Oxidation (E vs. NHE = 1.23 eV):  $H_2O + 2h^+ \rightarrow 2H^+ + \frac{1}{2}O_2$

Reduction (E vs. NHE = 0 eV):  $2H^+ + 2e^- \rightarrow H_2$

Overall process:  $H_2O \xrightarrow{(green-P)^*/h\upsilon} H_2 + \frac{1}{2}O_2$

The mechanism of water splitting involves the excitation of electrons from the valence band to conduction band of green phosphorene (monolayer and bilayer) by illuminating it with light of frequency greater than the bandgap. Then, the holes in the valence band combines with the adsorbed water molecules to form H$^+$ and O$_2$, if their energy is greater than the 1.23 eV (NHE).

The $H^+$ produced combines with the electrons in the conduction band to form a $H_2$ molecule. Thus, the water molecule in the presence of green phosphorene (in the excited state) gets reduced to hydrogen and oxygen molecules. This is a general mechanism of water splitting and has been explained in various studies by performing oxidation and redox reactions of water on the surface of the photocatalyst [39]. In this paper, however, we are mainly concentrated in exploring the possibility of modelling the band alignments and bandgap of green phosphorene (monolayer and bilayer) by strain-engineering or variation in the pH value, to make green phosphorene a suitable candidate for water splitting.

We found that the band gap of monolayer green phosphorene is more than 1.23 eV. The tuning of band gap by applying mechanical strains can also be achieved in experimental situation [40]. Our calculations show that the change in band gap with in-plane uniaxial tensile and compression strain is highly anisotropic, which is attribute to the structural anisotropy of green phosphorene [Figure S3]. We found that application of strain enables the bandgap of AC-stacked bilayer green phosphorene to be greater than 1.23 eV. The change in VBM and CBM on the application of strain is given in Figure 4. The maximum bandgap of 1.49 eV (1.37 eV) and 1.46 eV (1.36 eV) has been achieved in monolayer (AC-stacked bilayer) green phosphorene at uniaxial strain of 4% (4%) along x direction and -3% (-4%) along y direction, respectively.

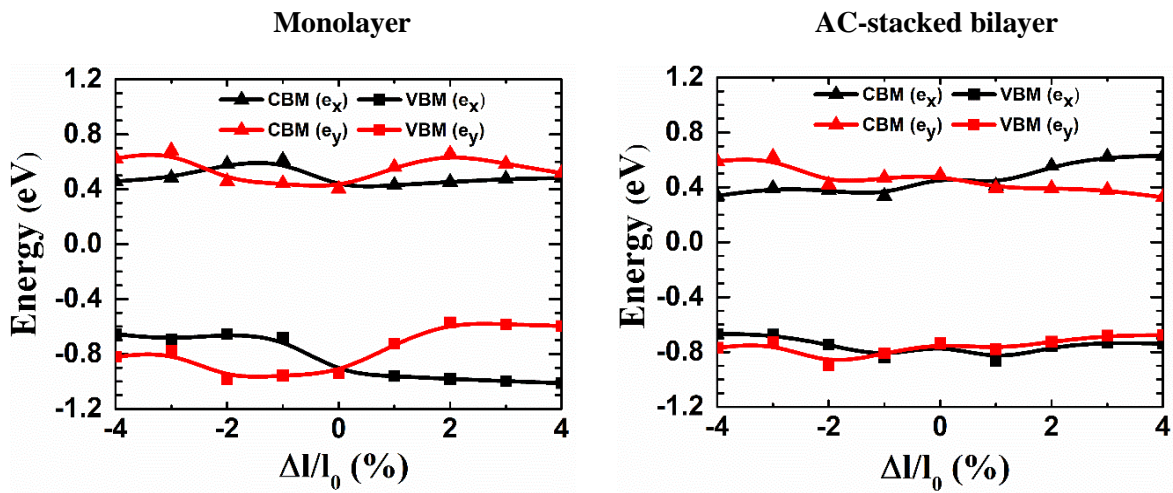

**FIGURE 4:** The conduction band minimum (CBM) and the valence band maximum (VBM) versus strain for monolayer and AC-stacked bilayer green phosphorene. Fermi energy plus vacuum energy are set to zero. For monolayer, the bandgap is highest at Δx/x = 4% (Bandgap = 1.49 eV), Δy/y = -3% (Bandgap = 1.46 eV) and for AC-stacked bilayer it is highest at Δx/x = 4% (Bandgap = 1.37 eV), Δy/y = -4% (Bandgap = 1.36 eV).

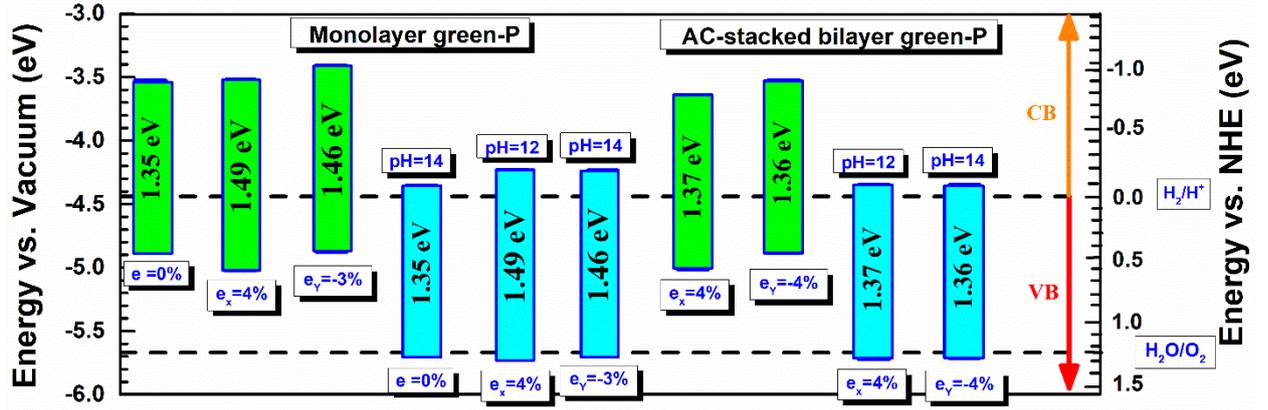

**FIGURE 5:** Energy alignment of monolayer and bilayer green phosphorene. The strain engineering and pH value of medium can modulate the band alignment. Green colored bars have zero pH value.

For photocatalytic water splitting, not only the band gap should be appropriate but the band alignments should also match with the redox potential of water. It is found that the VBM in monolayer and AC-stacked bilayer green-P that is less negative than the standard oxidation potential [Figure 5]. Therefore, as such, these systems cannot be used for water splitting. However, the standard redox potential depends on the pH of the solution. The standard oxidation potential $H_2O/O_2$ in a solution is given as [41]:

$$E^{ox}_{O_2/H_2O} = 5.67\ eV - pH \times 0.059\ eV$$

Hence the oxidation and redox potential will shift equally as the pH of the solution changes. We found that at pH =14, the monolayer depicts favorable band positions to be used as a photocatalyst. Similarly, for uniaxial strain of +4% along x direction and for uniaxial strain of -3% along y direction, where the bandgap reaches its maximum value, the favorable band alignment in monolayer for photocatalysis are obtained for pH=12 and pH=14, respectively.

Also, at pH value of 12 and 14, the band alignments of 4% (along x direction) and -4% (along y direction) uniaxialy strained AC stacked bilayer (i.e., the strain values where its bandgap value is maximum) becomes appropriate to be used as a catalyst for water splitting [Figure 5]. Thus, combining strain engineering with pH of solution, monolayer green-P and AC-stacked bilayer green-P can be potentially used for photocatalytic water splitting.

## 4. Green Phosphorene/MoSe$_2$ Heterostructure: Potential Application in Photovoltaics

To be used in applications such as photovoltaics, the materials should be capable of absorbing broad solar spectra and must possess high carrier mobility. Our calculations show that optical absorption for monolayer as well as bilayer AB and AC stacked green phosphorene, covers a wide range of solar spectrum i.e., from 1 eV to 4 eV which is significant for various applications based on solar energy [Figure S4].

In order to use bilayer green phosphorene in thin film photovoltaic system, an appropriate semiconducting acceptor materials required to be find out with matching band alignments. Specifically, it should form type II band alignments. Monolayer MoSe$_2$ has been found to exhibit the required band alignments for acceptor material. This direct nature of the bandgap [Figure 6] combined with high carrier mobility of monolayer and AB- and AC- stacked bilayer green phosphorene indicates their potential to be used as donor material for solar cell applications. The well-known that MoSe$_2$ monolayer possesses a sandwich type of structure with transition metal being sandwiched between the two layers of chalcogenides [35]. Note that the lattice mismatch between green phosphorene and monolayer MoSe$_2$ is less than 1% in both x- and y direction [Figure S5]. In the limit of 100% external quantum efficiency (EQE), the upper limit to the power conversion efficiency (PCE) is calculated as follows [42, 43]:

$$\eta = \frac{J_{sc}V_{oc}\beta_{FF}}{P_{solar}} = \frac{0.65(E_g^d - \Delta E_c - 0.3)\int_{E_g}^{\infty}\frac{P(\hbar\omega)}{\hbar\omega}d(\hbar\omega)}{\int_0^{\infty}P(\hbar\omega)d(\hbar\omega)}$$

where 0.65 is the band-fill factor (FF), $E_g^d$ is the bandgap of the donor, and the term ($E_g^d - \Delta E_c - 0.3$) estimates the maximum open circuit voltage $V_{oc}$. P($\hbar\omega$) is the solar energy flux at air mass 1.5 (AM1.5) at photon energy $\hbar\omega$. The integral in the numerator depicts the short circuit current $J_{sc}$ in the limit of 100% EQE while the denominator is the solar energy flux at AM1.5 [44].

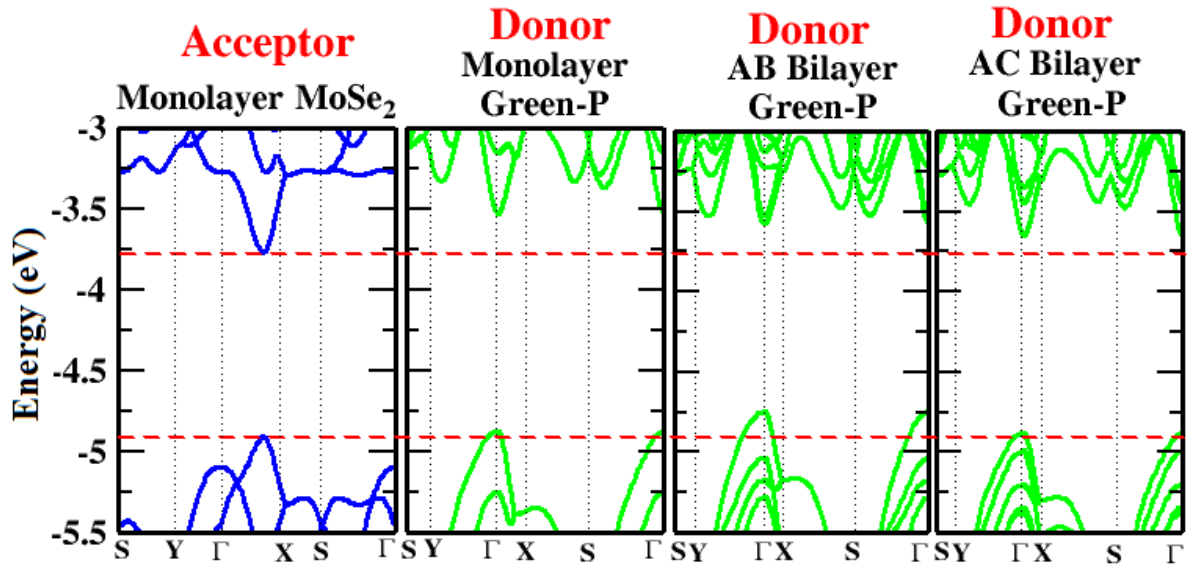

**FIGURE 6:** The electronic band structure of monolayer green phosphorene, AB- and AC-stacked bilayer green phosphorene and monolayer MoSe$_2$. Red dashed lines in the band structures is drawn to indicate the band alignment of green-phosphorene monolayer and bilayers w.r.t monolayer MoSe$_2$. The vacuum level is set to be zero.

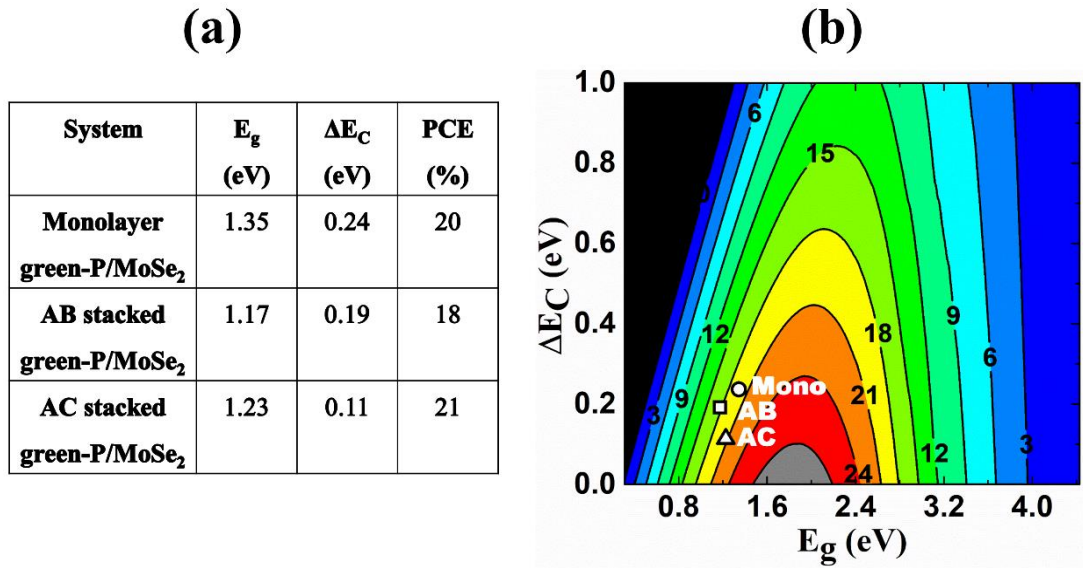

**FIGURE 7:** (a) Table for bandgap of the donor (E$_g$), conduction band offsets (ΔE$_c$) and percentage power conversion efficiencies (PEC) of the heterostructure of monolayer, AB- and AC-stacked bilayer green phosphorene with monolayer MoSe$_2$. (b) Calculated power-conversion efficiency contour as a function of donor bandgap (E$_g$) and conduction band offsets (ΔE$_c$) of monolayer AB- and AC-stacked bilayer green phosphorene with monolayer MoSe$_2$.

Our calculations show that the heterostructure constructed using monolayer, AB- and AC- stacked bilayer green phosphorene with MoSe$_2$ can achieve PCEs as high as 20%, 18% and 21%, respectively [Figure 7]. These values are comparable to those obtained for AA- and AB- stacked bilayer black phosphorene and MoS$_2$ (~18% and 16%) [24], α-AsP/GaN (~22.1%) [45] and PCBM/CBN systems (10-20%) [42]. Note that the PCE value depends on $E_g^d$ and $\Delta E_c$.

## 5. Summary

In summary, electronic properties including the carrier mobility of the monolayer and bilayer green phosphorene are reported using state-of-the-art density functional theory. Semiconductor-to-metal transition has been found to be induced in bilayer systems through electric field, vertical pressure and mechanical strain. Carrier mobility are calculated to be highly anisotropic and stacking-dependent. Monolayer and AB-stacked bilayer green phosphorene possess *n*-type semiconducting behaviour owing to their higher electron mobility. On the other hand, AC stacked bilayer exhibit *n*-type and *p*-type character along armchair and zigzag direction, respectively. The high carrier mobility and sizable band gap of green phosphorene are important for the photovoltaics and water splitting photocatalysis. It is found that the strained monolayer and bilayer green phosphorene can be potential candidates for solar water splitting catalysis in highly basic medium. The upper limit to the power conversion efficiencies of monolayer, AB- and AC-stacked bilayers heterostructure with $MoSe_2$, is calculated to be comparable with black phosphorene/$MoS_2$ based heterostructure. Hence, our results show the possibility of monolayer and bilayers of green phosphorene to be used in photovoltaic and photocatalytic applications.

## Acknowledgements

SK is grateful to UGC-BSR for financial assistance in the form of senior research fellowship. AK gratefully acknowledges University Grants Commission for Start-up grants (30-318/2016 (BSR)). The computational facility at Central University of Punjab, Bathinda and RAMA, High Performance Computing Cluster at Michigan Technological University Houghton, USA, are used to obtain the results presented in this paper.